\documentclass{emulateapj}
\usepackage{amsmath}
\usepackage{amssymb}
\usepackage{epsfig}
\usepackage{subfigure}
\usepackage{natbib}
\usepackage{graphicx}
\usepackage{color}

\shorttitle{The Type Ibn Supernova PS1-12sk}
\shortauthors{Sanders et al.}

\newcommand{\MJmass}{$2.62^{+0.56}_{-0.52} \times10^{11}~\rm{M}_{\odot}$}
\newcommand{\MJsfr}{$0.1^{+0.5}_{-0.1}~\rm{M}_{\odot}$~yr$^{-1}$}
\newcommand{\MJssfr}{$4.9^{+16.6}_{-2.9}\times10^{-4}~\rm{Gyr}^{-1}$}

\newcommand{\clusterM}{14.1_{-1.1}^{+1.1}}

\newcommand{\evlaepone}{+7}
\newcommand{\evlaeptwo}{+53}
\newcommand{\zrise}{9}
\newcommand{\zrisemax}{23}

\newcommand{\specepzero}{+6}
\newcommand{\specepone}{+35}
\newcommand{\speceptwo}{+10}
\newcommand{\specepthree}{+17}
\newcommand{\specepfour}{+8}

\newcommand{\pearsonfiveeightsevensix}{-0.22}
\newcommand{\pearsonsevenzerosixfive}{-0.66}

\newcommand{\allCdwarfs}{47}
\newcommand{\galfitblobg}{24.9\pm0.2}
\newcommand{\galfitblobMg}{-12.0\pm0.2}
\newcommand{\galfitblobSMCg}{60}
\newcommand{\galfitblobr}{24.3\pm0.1}
\newcommand{\galfitblobMr}{-12.6\pm0.1}
\newcommand{\galfitblobi}{24.3\pm0.1}
\newcommand{\galfitblobMi}{-12.6\pm0.1}
\newcommand{\galfitblobz}{24.7\pm0.2}
\newcommand{\galfitblobMz}{-12.2\pm0.2}

\newcommand{\perblueNtwo}{7}

\newcommand{\perblueNfour}{12}

\newcommand{\perblueNeight}{14}

\newcommand{\snlsRrblob}{5.6}
\newcommand{\snlsRrcgc}{3.1}

\newcommand{\DL}{238}
\newcommand{\distmod}{36.88}

\newcommand{\MMTsfr}{2\times10^{-3}}
\newcommand{\MMTHaL}{2\times10^{38}}

\newcommand{\radRise}{(8-20)\times10^{14}}
\newcommand{\tEject}{1.8-4.5}

\newcommand{\PeakjcB}{-17.9}
\newcommand{\PeakjcV}{-17.7}
\newcommand{\PeakjcR}{-17.6}
\newcommand{\PeakjcI}{-17.6}

\newcommand{\MNiCoOneT}{59}
\newcommand{\MNiCoOne}{2.0}
\newcommand{\MNiCoOneGT}{1.5}

\newcommand{\MNiCoTwo}{2.4}
\newcommand{\MNiCoTwoGT}{1.7}

\newcommand{\PolyPeakgM}{-19.24}
\newcommand{\PolyPeakgMerr}{0.02}

\newcommand{\bolomaxt}{49}
\newcommand{\bolomint}{-10}
\newcommand{\boloL}{6.4\pm0.4}
\newcommand{\boloLHeM}{0.06}
\newcommand{\boloLHeMdot}{0.01}
\newcommand{\boloLHeMten}{0.002}
\newcommand{\boloMag}{-4.33\pm0.06}
\newcommand{\gmrmederr}{0.1}
\newcommand{\peakBBrad}{1.4}
\newcommand{\peakBBradD}{0.3}

\newcommand{\PolyPeakrM}{-19.08}
\newcommand{\PolyPeakrMerr}{0.02}
\newcommand{\PolyPeakrdmfift}{1.44}
\newcommand{\PolyPeakrdmfifterr}{0.07}

\newcommand{\PolyPeakiM}{-18.44}
\newcommand{\PolyPeakiMerr}{0.03}

\newcommand{\PolyPeakzt}{56006.1}
\newcommand{\PolyPeakzterr}{0.3}
\newcommand{\PolyPeakzM}{-18.88}
\newcommand{\PolyPeakzMerr}{0.02}

\newcommand{\skMR}{-19.6}

\newcommand{\skMNione}{0.5}
\newcommand{\skMejone}{0.3}
\newcommand{\skEKone}{0.2}

\newcommand{\gps}{\ensuremath{g_{\rm P1}}}
\newcommand{\rps}{\ensuremath{r_{\rm P1}}}
\newcommand{\ips}{\ensuremath{i_{\rm P1}}}
\newcommand{\zps}{\ensuremath{z_{\rm P1}}}
\newcommand{\yps}{\ensuremath{y_{\rm P1}}}

\newcommand{\grizy}{\gps\rps\ips\zps\yps}
\newcommand{\PS}{\protect \hbox {Pan-STARRS1}}
\def\amin{\char'023 }
\def\asec{\char'175 }

\newcommand{\kms}{{\rm km~s}^{-1}}

\newcommand{\aCfA}{1}
\newcommand{\aQUB}{2}
\newcommand{\aASIAA}{3}
\newcommand{\aNCU}{4}
\newcommand{\aCarn}{5}
\newcommand{\aSTSI}{6}
\newcommand{\aTexT}{7}
\newcommand{\aMSU}{8}
\newcommand{\aMIT}{9}
\newcommand{\aIfA}{10}
\newcommand{\aPrince}{11}

\begin{document}

\title{PS1-12sk is a Peculiar Supernova From a He-rich Progenitor System\\ in a Brightest Cluster Galaxy Environment}
\author{
N.~E.~Sanders,\altaffilmark{\aCfA}
A.~M.~Soderberg,\altaffilmark{\aCfA}
R.~J.~Foley,\altaffilmark{\aCfA}
R.~Chornock,\altaffilmark{\aCfA}
D.~Milisavljevic,\altaffilmark{\aCfA}
R.~Margutti,\altaffilmark{\aCfA}
M.~R.~Drout, \altaffilmark{\aCfA}
M.~Moe,\altaffilmark{\aCfA}
E.~Berger,\altaffilmark{\aCfA}
W.~R.~Brown,\altaffilmark{\aCfA}
R.~Lunnan,\altaffilmark{\aCfA}
S.~J.~Smartt,\altaffilmark{\aQUB}
M.~Fraser,\altaffilmark{\aQUB}
R.~Kotak,\altaffilmark{\aQUB}
L.~Magill,\altaffilmark{\aQUB}
K.~W.~Smith,\altaffilmark{\aQUB}
D.~Wright,\altaffilmark{\aQUB}
K.~Huang,\altaffilmark{\aASIAA}
Y.~Urata,\altaffilmark{\aNCU}
J.~S.Mulchaey,\altaffilmark{\aCarn}
A.~Rest,\altaffilmark{\aSTSI}
D.~J.~Sand,\altaffilmark{\aTexT}
L.~Chomiuk,\altaffilmark{\aMSU}
A.~S.~Friedman,\altaffilmark{\aCfA}$^,$\altaffilmark{\aMIT}
R.~P.~Kirshner,\altaffilmark{\aCfA}
G.~H.~Marion,\altaffilmark{\aCfA}
J.~L.~Tonry, \altaffilmark{\aIfA}
W.~S.~Burgett,\altaffilmark{\aIfA}
K.~C.~Chambers,\altaffilmark{\aIfA} 
K.~W.~Hodapp,\altaffilmark{\aIfA}
R.~P.~Kudritzki,\altaffilmark{\aIfA}
P.~A.~Price\altaffilmark{\aPrince}
} 

\altaffiltext{\aCfA}{Harvard-Smithsonian Center for Astrophysics, 60 Garden Street, Cambridge, MA 02138 USA}
\altaffiltext{\aQUB}{Astrophysics Research Centre, School of Maths and Physics,Queen’s University, BT7 1NN, Belfast, UK}
\altaffiltext{\aASIAA}{Academia Sinica Institute of Astronomy and Astrophysics, Taipei 106, Taiwan}
\altaffiltext{\aNCU}{Institute of Astronomy, National Central University, Chung-Li 32054, Taiwan}
\altaffiltext{\aCarn}{Carnegie Observatories, 813 Santa Barbara Street, Pasadena, CA 91101, USA}
\altaffiltext{\aSTSI}{Space Telescope Science Institute, 3700 San Martin Dr., Baltimore, MD 21218, USA}
\altaffiltext{\aTexT}{Texas Tech University, Physics Department, Box 41051, Lubbock, TX 79409-1051}
\altaffiltext{\aMSU}{Department of Physics and Astronomy, Michigan State University, East Lansing, Michigan 48824, USA}
\altaffiltext{\aMIT}{Massachusetts Institute of Technology, 77 Massachusetts Ave., Bldg. E51-173, Cambridge, MA 02138, USA}

\altaffiltext{\aIfA}{Institute for Astronomy, University of Hawaii, 2680 Woodlawn Drive, Honolulu HI 96822}
\altaffiltext{\aPrince}{Department of Astrophysical Sciences, Princeton University, Princeton, NJ 08544, USA}

\email{nsanders@cfa.harvard.edu}

\begin{abstract}

We report on our discovery and observations of the \PS\ supernova (SN) PS1-12sk, a transient with properties that indicate atypical star formation in its host galaxy cluster or pose a challenge to popular progenitor system models for this class of explosion.  The optical spectra of PS1-12sk classify it as a Type~Ibn SN (c.f. SN~2006jc), dominated by intermediate-width ($3\times10^3~\kms$) and time variable \ion{He}{1} emission.  
Our multi-wavelength monitoring establishes the rise time $dt\sim9-\zrisemax$~days and shows an NUV-NIR SED with temperature $\gtrsim17\times10^3$~K and a peak magnitude of $M_z=\PolyPeakzM\pm\PolyPeakzMerr$~mag.  
SN~Ibn spectroscopic properties are commonly interpreted as the signature of a massive star ($17-100~\rm{M}_\odot$) explosion within a He-enriched circumstellar medium.  However,  unlike previous Type Ibn supernovae, PS1-12sk is associated with an elliptical brightest cluster galaxy, CGCG~208-042 ($z=0.054$) in cluster RXC~J0844.9+4258.  
The expected probability of an event like PS1-12sk in such environments is low given the measured infrequency of core-collapse SNe in red sequence galaxies compounded by the low volumetric rate of SN~Ibn.   
Furthermore, we find no evidence of star formation at the explosion site to sensitive limits ($\Sigma_{\rm{H}\alpha} \lesssim  2\times 10^{-3}~{\rm M}_{\odot}~{\rm yr}^{-1}~\rm{kpc}^{-2}$).   
We therefore discuss white dwarf binary systems as a possible progenitor channel for SNe~Ibn. We conclude that PS1-12sk represents either a fortuitous and statistically unlikely discovery, evidence for a top-heavy IMF in galaxy cluster cooling flow filaments, or the first clue suggesting an alternate progenitor channel for Type~Ibn SNe.

\smallskip
\end{abstract}


\section{INTRODUCTION}
\label{sec:intro} 

Traditionally, hydrogen-poor supernovae (Type~I SNe) have been classified into three sub-classes based on the presence of Si (Type~Ia), He (Type~Ib), or the absence of both features (Type~Ic) in their optical spectra (see \citealt{Filippenko97} for a review).  Since the discovery of SN~1999cq \citep{Matheson00}, a new sub-class of ``Type~Ibn'' SNe have emerged, characterized by intermediate-width (${\rm FWHM}\sim3\times10^3~\kms$) \ion{He}{1} emission.  The most well studied of these SNe~Ibn is SN~2006jc \citep{Pastorello07}, while other examples identified in the literature are limited to 2000er \citep{Pastorello08}, 2002ao \citep{Foley07}, 2011hw \citep{Smith12}, and perhaps SN~2005la \citep{Pastorello05la}.

Several lines of evidence point to a massive star ($\sim17-100~\rm{M}_\odot$) origin for SNe~Ibn.  First, the \ion{He}{1} emission is representative of a dense circumstellar medium (CSM), suggesting a progenitor with an He-rich envelope and high mass loss rate, such as a Wolf Rayet star (\citealt{Foley07,Pastorello08,Smith08,Tominaga08}).  Second, a Luminous Blue Variable (LBV)-like eruption was observed at the location of SN~2006jc $\sim2$~yr before the SN explosion \citep{Pastorello07}. Third, late-time ($\sim2$~months) IR and spectroscopic observations of SN\,2006jc suggest hot carbon dust formation in the SN ejecta, with total $M_{ej}\sim5~\rm{M}_\odot$ \citep{DiCarlo08,Mattila08,Nozawa08,Smith08,Tominaga08,Sakon09}.  Fourth, all past SNe~Ibn have been found in star-forming galaxies, consistent with a massive star progenitor.   Intermediate-width H emission has been detected in the spectra of some SNe~Ibn, with strengths significantly weaker than the \ion{He}{1} lines \citep{Pastorello08,Smith08,Smith12}
.
Intermediate-width H$\alpha$ emission suggests a connection between SNe~Ibn and IIn -- a class whose spectra are dominated by intermediate-width H spectral features and are in some cases associated with LBV-like progenitors \citep{Pastorello05la,GalYam07,GalYam05gl,Smith10jl,Kochanek11,Mauerhan12}.  The close temporal connection between these LBV-like events and the SN explosions suggests a massive star progenitor, but also challenges models for massive star evolution that predict stars should spend the final $\sim1$~Myr of their lives in a core-He burning Wolf-Rayet phase \citep{Heger03,Smith12}.  

Here we present observations of a newly discovered SN~Ibn found in a host environment with no direct evidence of a young stellar population.  The optical transient PS1-12sk was discovered on 2012~March~11 by the Panoramic Survey Telescope \& Rapid Response System 1 survey (\PS, abbreviated PS1, \citealt{PS1}) at $\zps=18.66\pm0.01$~mag at position $08^{\rm h}44^{\rm m}54.86^{\rm s}~+42^{\circ}58\amin16.89\asec$ (J2000), within the galaxy cluster RXC~J0844.9+4258.  The object was spectroscopically classified as the first Type~Ibn SN discovered by \PS\ after just $\sim2$~years of survey operation.  At $z=0.054$, PS1-12sk is more distant than any previously discovered SN~Ibn.\footnote{We assume a standard $\Lambda$CDM cosmology throughout this work, adopting the Hubble constant $H_0=71~\kms~\rm{Mpc}^{-1}$, a distance modulus for PS1-12sk and its host environment (CGCG~208-042) of $\mu=\distmod$, and a luminosity distance of \DL~Mpc.}

We describe our multi-wavelength (radio through X-ray) observations of PS1-12sk in Section~\ref{sec:obs}.  In Section~\ref{sec:comp}, we discuss the observed properties of PS1-12sk and compare to past SNe~Ibn.  Our multi-wavelength monitoring of PS1-12sk provides the most detailed information to date on the rise phase and NUV-NIR SED of a Type~Ibn SN.  Deep stacks of pre-explosion PS1 imaging and optical spectroscopy allow us to characterize the host environment of PS1-12sk in depth (Section~\ref{sec:res:host}), pointing to the massive elliptical brightest cluster galaxy CGCG~208-042 as the most likely host galaxy and placing strong limits on star formation levels at the explosion site.  
We infer characteristics of the progenitor system from our observations of PS1-12sk in Section~\ref{sec:disc}.  In light of the observed explosion and host environment properties, we discuss several possible massive star and white dwarf progenitor channels for this SN in Section~\ref{sec:prog}, and speculate on their implications for the initial mass function (IMF) and star formation in the cluster environment.  We conclude in Section~\ref{sec:conc}.

\section{OBSERVATIONS}
\label{sec:obs}

\subsection{Optical Photometry}
\label{sec:obs:opt}

After discovery with PS1 on MJD~55997, we monitored the optical evolution of PS1-12sk through MJD~56049 (see Figure~\ref{fig:LC}), after which it went into conjunction with the sun.  These observations are summarized in Table~\ref{tab:LC} and described below.

PS1 is a high-etendue wide-field imaging system, designed for dedicated survey observations. The system is installed on the peak of Haleakala on the island of Maui in the Hawaiian island chain. Routine observations are conducted remotely, from the University of Hawaii--Institute for Astronomy Advanced Technology Research Center (ATRC) in Pukalani. A complete description of the PS1 system, both hardware and software, is provided by \cite{PS1}. The PS1 optical design is described in \cite{PS1opt}, the imager is described in \cite{PS1cam}, and the survey design and execution strategy are described in \cite{PS_MDRM}.  The standard reduction, astrometric solution, and stacking of the nightly images is done by the Pan-STARRS1 IPP system \citep{PS1_IPP,PS1_astrometry}. The nightly Medium Deep stacks are transferred to the Harvard FAS Research Computing cluster, where they are processed through a frame subtraction analysis using the \textit{photpipe} image differencing pipeline developed for the 
SuperMACHO and ESSENCE surveys \citep{Rest05,Garg07,Miknaitis07}.  A summary of details of PS1 operations relevant to SN studies is given in \cite{Chomiuk11}, and additional PS1 SN studies were presented by \cite{Gezari10,Botticella10,Chomiuk11,Narayan11,Berger12,nes2010ay,Valenti12,Chornock13,Lunnan13}.

The PS1 observations are obtained through a set of five broadband filters, which we have designated as \gps, \rps, \ips, \zps, and \yps \citep{PS1cal}. Although the filter system for PS1 has much in common with that used in previous surveys, such as the Sloan Digital Sky Survey \citep[SDSS:][]{York00,SDSS8}, there are important differences. The \gps\ filter extends $200~{\rm \AA}$ redward of $g_{\rm SDSS}$, and the \zps\ filter is cut off at $9200~{\rm \AA}$. SDSS has no corresponding \yps\ filter. Further information on the passband shapes is described in \cite{PS1cal}. Photometry is in the ``natural'' PS1 system, $m = −2.5 \log(\rm flux) + m^\prime$, with a single zero-point adjustment $m^\prime$ made in each band to conform to the AB magnitude scale \citep{Tonry12}.  We assume a systematic uncertainty of 1\% for our PS1 observations due to the asymmetric PS1 point spread function and uncertainty in the photometric zero-point calibration \citep{Tonry12}.

PS1 observed the field of PS1-12sk in two prior seasons dating to December 2009 (MJD~55174), but we detect no transient flux before the March 2012 explosion.  Considering observations in all filters, the $[2009-2010, 2010-2011,2011-2012]$ pre-explosion observing seasons had durations of $[10,27,15]$~weeks and typical cadences of $[4,4,1]$~days, respectively.  The pre-explosion imaging consists of $[16,18,22,32,21]$ frames in $[\gps,\rps,\ips,\zps,\yps]$~band with median $3\sigma$ limiting absolute magnitude of $M_{g,r,i,z,y}\gtrsim[-14.1,-13.8,-15.4,-14.6,-15.0]$~mag, respectively.  These pre-explosion limits are not strongly constraining in the context of a SN~2006jc-like pre-explosion outburst, an LBV-like flare with peak magnitude $M_r\sim-14.1$~mag that was detected for just nine days \citep{Pastorello07}.  

\begin{figure*}
\includegraphics[width=7in]{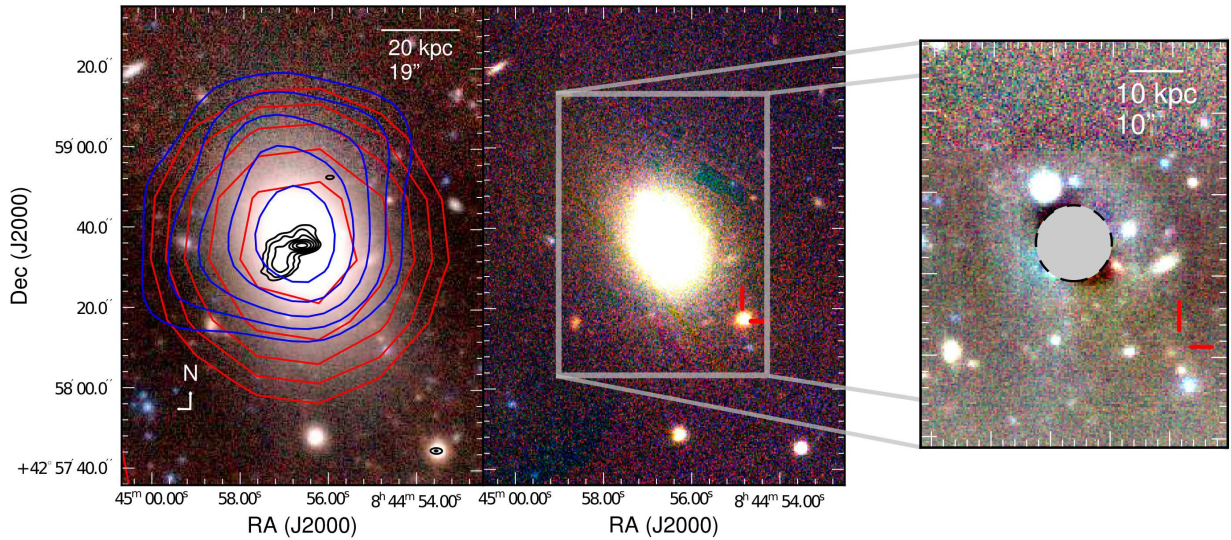}
\caption{\label{fig:hostcomposite}\label{fig:stacks:g:rgb}Multiwavelength observations of the field of PS1-12sk.  Left: optical composite of $\gps\rps\zps$ pre-explosion PS1 images from December 2009 to May 2012, shown with log stretch.  Contours from our 5.9~GHz JVLA observations are overlaid in black (beam $\rm{FWHM}\sim3.5$\asec), NVSS 1.4~GHz radio observations are overlaid in red (beam $\rm{FWHM}\sim45$\asec), and XMM-Newton X-ray observations are overlayed in blue ($\rm{FWHM}\sim6$\asec).  The radio and X-ray emission is centered on CGCG~208-042, the brightest cluster galaxy.  Middle: PS1 $\gps\rps\zps$  imaging of the environment near the peak of PS1-12sk (2012 March 13; MJD~55999), shown with linear stretch.  The SN position is marked.  Right: The deep composite with the model for the BCG subtracted (linear stretch; a $1\times1.33$\arcmin\ field).  The circular mask excludes the inner 7\asec of the residual image for source detection.}
\end{figure*}

Additional $ugriz$ imaging was acquired with the 2.0~m Liverpool Telescope with the optical CCD camera RATCam from MJD~$56015-56039$.  RATCam data were reduced following \cite{Valenti11}.  Fixed aperture ($3\asec$) photometry of PS1-12sk was performed using SExtractor \citep{sextractor} and zero points were measured from comparison with field stars in the SDSS catalog.  Flux from CGCG~208-042 at the position of PS1-12sk was below the noise level of the RATcam observations, so template subtraction was not performed.

UV/optical observations of PS1-12sk with the Swift-UVOT instrument \citep{Roming05} were acquired from MJD~$56013-56032$.  Observations were performed using all of its six broad band filters, spanning the wave-length range $\lambda_c=1928$ Angstroms ($W2$ filter) - $\lambda_c=5468$ Angstroms ($V$ filter, central wavelength). Starting from MJD~56041 the SN was too faint in the UV wavelength range and we therefore limited our follow up to the $U$, $B$ and $V$ filters. Data have been analyzed following the prescriptions by \cite{Brown09}. In particular: a  $3\asec$ aperture has been used to maximize the signal-to-noise ratio. The AB photometry presented in Table~\ref{tab:LC} is based on the UVOT photometric system of \cite{Poole08}.  Host galaxy imaging was not available for template subtraction, but the host galaxy flux at the explosion site is not significant compared to the uncertainty in the SN flux.

\begin{deluxetable}{lrll}
\tablecaption{PS1-12sk Optical/NIR Photometry\label{tab:LC}}
\tablehead{ \colhead{MJD} & \colhead{Filter} & \colhead{$m$ (AB mag)} & \colhead{Instrument} }
\scriptsize
\startdata
56021.3	&	H	&	$21.24\pm0.06$	&	CFHT\\
56021.3	&	J	&	$20.29\pm0.08$	&	CFHT\\
56021.3	&	K	&	$21.80\pm0.06$	&	CFHT\\
56028.2	&	H	&	$21.87\pm0.08$	&	CFHT\\
56028.2	&	J	&	$21.02\pm0.07$	&	CFHT\\
56028.2	&	K	&	$22.21\pm0.06$	&	CFHT\\
56013.1	&	H	&	$>19.27$	&	PAIRITEL\\
56013.1	&	J	&	$18.86\pm0.10$	&	PAIRITEL\\
56013.1	&	K	&	$>18.35$	&	PAIRITEL\\
56014.2	&	H	&	$>19.08$	&	PAIRITEL\\
\enddata
\tablecomments{Uncertainties reflect $1\sigma$ ranges while upper limits are $3\sigma$.  Table~\ref{tab:LC} is published in its entirety in the electronic edition.  A portion is shown here for guidance regarding its form and content.}
\end{deluxetable}

\subsection{Optical spectroscopy}
\label{sec:obs:spectra}

\begin{deluxetable}{llllr}
\tablecaption{PS1-12sk Optical Spectroscopy\label{tab:spec}}
\tablehead{
\colhead{MJD} & \colhead{Telescope} & \colhead{Epoch\tablenotemark{a}} & \colhead{Range}  & \colhead{Dispersion} \\
\colhead{} & \colhead{/Instrument} & \colhead{(Days)} & \colhead{(\AA)}  & \colhead{($\rm{\AA~px}^{-1}$)}
}
\scriptsize
\startdata
56012.5 & MMT/BC & \specepzero & $3330-8530$ & 1.9 \\          
56014.2 & MMT/BC & \specepfour & $5870-7770$ & 0.7 \\          
56016.5 & NOT/AFOSC & \speceptwo & $3520-8980$ & 3.0 \\        
56023.5 & INT/IDS & \specepthree & $3695-9485$ & 4.1 \\        
56041.5 & MMT/BC & \specepone & $3535-8410$ & 1.9              
\enddata
\tablenotetext{a}{Epoch relative to $z$-band peak.}
\end{deluxetable}

\begin{figure*}
\plotone{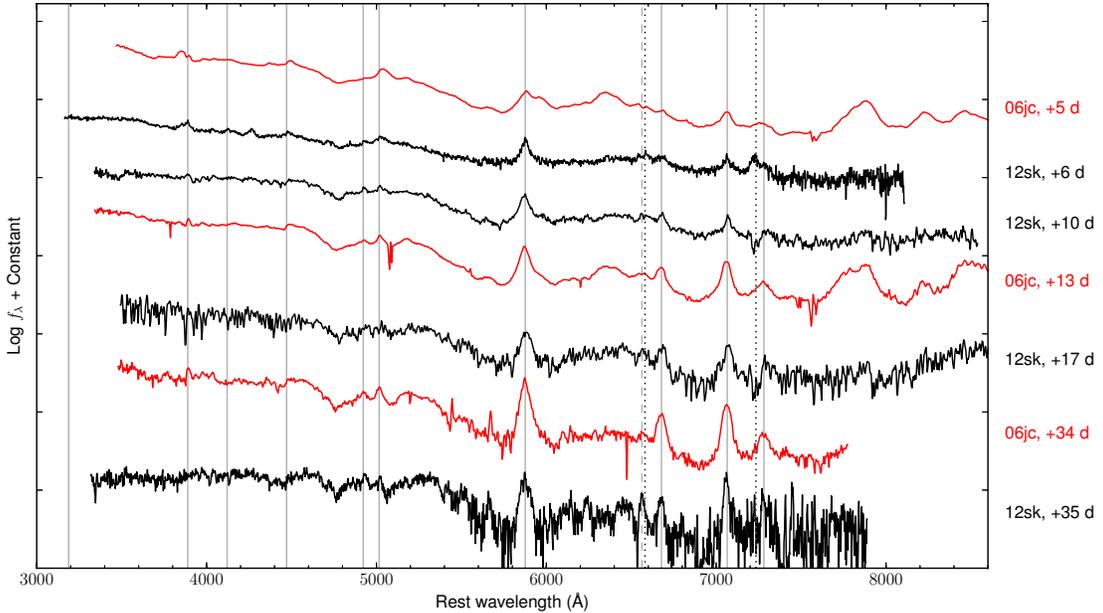}
\caption{\label{fig:spectra}Low-resolution spectroscopic sequence of PS1-12sk, with time since $z$-band peak noted at right.  See Table~\ref{tab:spec} for observing details.  The moderate resolution MMT/BC spectrum is shown separately in Figure~\ref{fig:specHR}.  The locations of the \ion{He}{1} features $\lambda\lambda3188,3889,4121,4471,4922,5016,5876,6678,7065,7281$~\AA\ are marked with solid lines; H$\alpha$ with a dashed line; and \ion{C}{2}~$\lambda\lambda6580,7234$ with dotted lines.  Spectra of SN\,2006jc at representative epochs \citep{Pastorello08} are shown in red.}
\end{figure*}

Our spectroscopic observations are summarized in Table~\ref{tab:spec} and displayed in Figure~\ref{fig:spectra}.   Spectroscopy of PS1-12sk was first obtained with the BlueChannel (BC) spectrograph of the MMT \citep{Schmidt89} on MJD~56012.5.  A moderate resolution ($0.7\rm{\AA~px}^{-1}$) BC spectrum was taken on MJD~56014.2.  Subsequent spectroscopy was performed with the Andalucia Faint Object Spectrograph and Camera of the Nordic Optical Telescope on MJD~56016.5, the Intermediate Dispersion Spectrograph of the Isaac Newton Telescope on MJD~56023.5, and BC on MJD~56041.5.  All spectra were reduced using standard two-dimensional long-slit image reduction and spectral extraction routines in IRAF\footnote{IRAF is distributed by the National Optical Astronomy Observatory, which is operated by the Association of Universities for Research in Astronomy (AURA) under cooperative agreement with the National Science Foundation.} and flux 
calibrated using 
observations of spectrophotometric standard stars obtained on the same night. 

\subsection{NIR imaging}
\label{sec:obs:pairitel}

We obtained NIR photometry of PS1-12sk using simultaneous $JHK$ imaging from the robotic 1.3~m Peters Automated Infrared Imaging Telescope
(PAIRITEL; \citealt{BloomP}) at Mount Hopkins, Arizona in 18~epochs from MJD~$56013-56034$.  PAIRITEL images were reduced using the CfA pipeline described in \cite{WoodVasey08} and \cite{Mandel09}, with more detailed discussion of the updated mosaic and photometry pipelines to be discussed in Friedman et al. (2013, in prep.).  We did not perform template subtraction to remove the host flux, as the NIR flux from CGCG~208-042 at the position of PS1-12sk was below the noise level of the PAIRITEL observations.  We use the 2MASS point source catalog to establish the photometric zero points \citep{Cutri03}.  

We obtained further imaging of PS1-12sk with WIRCAM \citep{WIRCAM} on the 3.6~m Canada-France-Hawaii Telescope (CFHT). NIR $J$, $H$ and $Ks$-band observations were carried out on MJD~56021 and 56028. The data were processed with the standard WIRCAM pipeline.  The photometric calibrations were performed against with the 2MASS point source catalog.

The NIR photometry from both instruments is listed in Table~\ref{tab:LC}.

\subsection{Radio Observations}
\label{sec:obs:evla}

We observed PS1-12sk with the Karl G. Jansky Very large Array (JVLA; \citealt{EVLA}) on two epochs, MJD~56013.0 and 56059.0 ([\evlaepone,\evlaeptwo]~d after $z$-band peak). All JVLA observations were obtained with two 1~GHz sidebands centered at 5.0 and 6.75~GHz, averaging to 5.9~GHz.  We used calibrator~J0818+4222 to monitor the phase and 3C147 for flux calibration. Data were reduced using the standard packages of the Astronomical Image Processing System (AIPS). 

We do not detect a radio counterpart to PS1-12sk in these observations (Figure~\ref{fig:stacks:g:rgb}) and place upper limits of $F_\nu \lesssim [16, 22]~\mu$Jy ($3\sigma$) for each epoch respectively, corresponding to upper limits on the spectral luminosity of $L_{\nu}\lesssim [1.1,1.5]\times 10^{27}~\rm{erg~s}^{-1}$. The SN~2006jc radio peak was 415 $\mu$Jy at 8.46 GHz (VLA program AS887, PI Soderberg; see also \citealt{Soderberg06jc}). Given the distance to SN~2006jc of 28~Mpc the luminosity is $3.9\times10^{26}~\rm{erg}~\rm{s}^{-1}~\rm{Hz}^{-1}$ at $\sim80$ days after explosion, a factor of $\sim3$ below our upper limits for PS1-12sk.  Figure~\ref{fig:stacks:g:rgb} illustrates that the 5.9~GHz morphology of the host galaxy cluster emission is asymmetric and peaked at the center of brightest cluster galaxy CGCG~208-042, suggesting evidence of an active nucleus.  Optical line emissions ratios verify the presence of AGN activity (Section~\ref{sec:res:host}).

The host environment of PS1-12sk is included in the NRAO VLA Sky Survey (NVSS) 1.4~GHz continuum survey \citep{Condon98}.  These observations are displayed in Figure~\ref{fig:hostcomposite}, which shows that the 1.4~GHz flux of $16.4\pm0.6$~mJy ($1.1\times10^{23}
~\rm{W~Hz^{-1}}$ at $z=0.054$) listed in the NVSS catalog is centered at the core of CGCG~208-042.

\subsection{X-ray Observations}
\label{sec:obs:xray}

We observed PS1-12sk with the X-Ray Telescope (XRT, \citealt{Burrows05}) onboard the \emph{Swift} spacecraft \citep{Gehrels04}, in the time period MJD~$56013-56056$, for a total of  26.4 ks. \emph{Swift}-XRT  data have been analyzed using the latest version of the HEASOFT package available at the time of writing (v.~6.12) and corresponding calibration files. Standard filtering and screening criteria have been applied. No point-like X-ray source is detected at the SN position. However, we find evidence for spatially extended X-ray emission from RXCJ0844.9+4258, the host galaxy cluster, at the SN location at the level of $(3\pm1)\times 10^{-14}\rm{erg\,s^{-1}sm^{-2}}$, corresponding to a luminosity of $(2.0\pm 0.7)\times 10^{41}\rm{erg\,s^{-1}}$. (0.3-10 keV). We assumed a simple power-law spectral model corrected for Galactic absorption in the direction of the SN with neutral H column density $N_H=2.93\times 10^{20}\rm{cm^{-2}}$ \citep{Kalberla05}.  Assuming the same spectral model, at the cluster core we 
measure a similar flux, $(4\pm1)\times10^{-14}~\rm{erg~s}^{-1}~\rm{cm}^{-2}$ (0.3-10 keV).  For comparison, \cite{Immler08} detected SN~2006jc with the Swift XRT at several epochs, finding a luminosity of $\sim1\times10^{39}~\rm{erg~s}^{-1}$ $\sim1$~month after explosion, rising to $\sim4$ times that level at $\sim4$~months.  The luminosity of SN~2006jc at any epoch was therefore well below the cluster background level we observe at the position of PS1-12sk.

We analyzed archival XMM-Newton data of the host galaxy cluster from 2007~November~6 (Program~50360; PI Mulchaey). The 26.8~ks observation was significantly impacted by high background levels due to solar flares.   Following the procedure outlined in \cite{Jeltema06}, but using version 12.0.0 of the SAS reduction software, we identified periods of high flaring and removed these times from the  dataset.   The final effective exposure times are approximately 23~ksec for the MOS detectors and 13 ksec for the PN detector. Within the usable portion of the dataset, we find $\sim 200$~photons associated with the cluster.   
The cluster X-ray emission is extended and detected out to a radius $\sim40\asec$ that includes the explosion site of PS1-12sk. The morphology of the X-ray emission shows that the emission is peaked at the center of the core of CGCG~208-042 as shown in Figure\ref{fig:hostcomposite}.

\section{Comparison to Past Type Ibn Supernovae}
\label{sec:comp}

\subsection{Light curve evolution}
\label{sec:res:lc}

\begin{figure}
\plotone{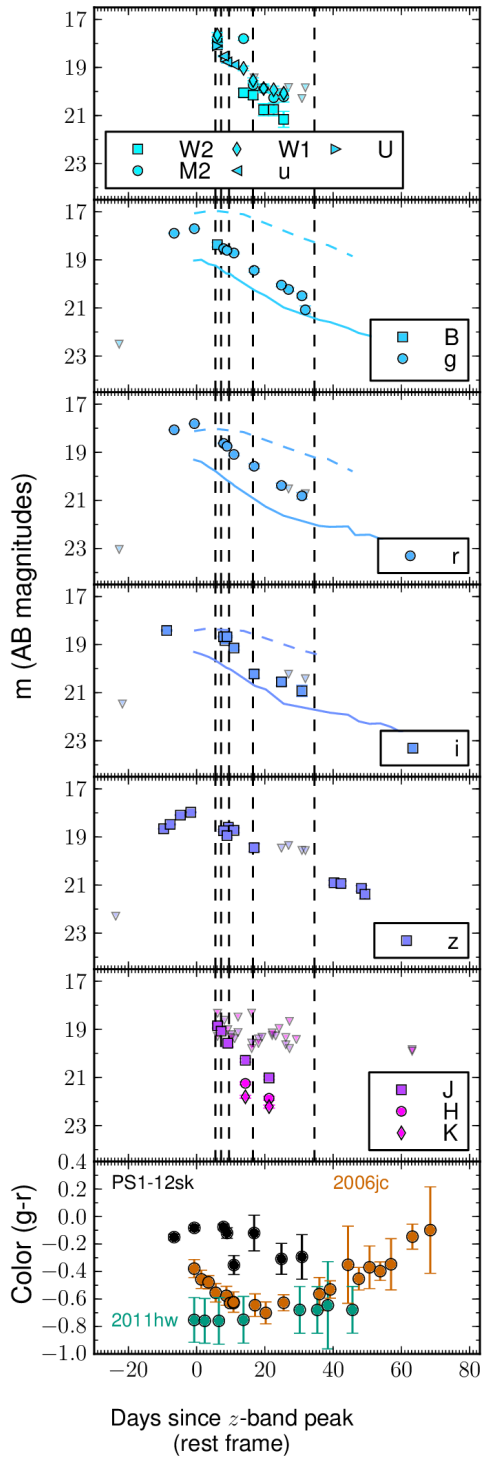}
\caption{\label{fig:LC}The multi-band light curve of PS1-12sk and $3\sigma$ upper limits (triangles).  The thin lines are the $BVRI$ light curves (AB magnitudes) of SN\,2006jc \citep[solid,][]{Foley07} and SN~2011hw \citep[dashed,][]{Smith12}, shifted temporally so the brightest/first observation corresponds to the epoch of $z$-band peak for PS1-12sk ($\rm{MJD}=\PolyPeakzt\pm\PolyPeakzterr$) and rescaled to $z=0.054$.  The bottom panel shows the optical color (AB magnitudes), corrected for Galactic extinction, of PS1-12sk, SN~2006jc, and SN~2011hw \citep{Smith12}.  No extinction or K-corrections have been applied.  The dashed vertical lines indicate the epochs of our optical spectroscopy.}
\end{figure}

We present the full, multi-band light curve constructed from our observations of PS1-12sk in Figure~\ref{fig:LC}.  During the course of PS1 $z$-band observations, PS1-12sk rose by $\sim1$~mag over $\sim9$~days and then faded by $\sim3$~mag over $\sim50$~days.  Using a fifth-order polynomial fit, we estimate the peak magnitude in $z$-band to be $\PolyPeakzM\pm\PolyPeakzMerr$~mag at $\rm{MJD}=\PolyPeakzt\pm\PolyPeakzterr$.  Similarly, in $[g,r,i]$-band we find $M_{\rm{peak}}=[\PolyPeakgM\pm\PolyPeakgMerr,\PolyPeakrM\pm\PolyPeakrMerr,\PolyPeakiM\pm\PolyPeakiMerr]$~mag.

\begin{figure}
\plotone{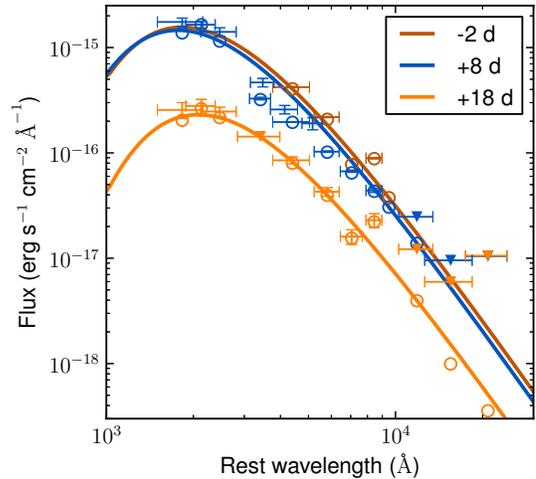}
\caption{\label{fig:BBpeak}The SED of PS1-12sk from photometry at three epochs (${\rm MJD}=[56005,56015,56025]\pm1.5$).  The circles show the linearly-interpolated light curve value in each filter at each precise epoch, error bars show the flux and filter width for the measured NUV-NIR photometric points nearest to each epoch (only when within 1.5 days), and triangles show the nearest $3\sigma$ upper limit.  The solid-lines show the blackbody fit to the interpolated photometry.  The NIR observations do not suggest a significant infrared excess for PS1-12sk within a few weeks of peak brightness.}
\end{figure}

We estimate the foreground extinction affecting PS1-12sk to be insignificant. The Galactic extinction at its position is $E(B-V)=0.03$~mag \citep{SF11} and the host galaxy is unlikely to have significant local dust obscuration (see Section~\ref{sec:res:host}).  Furthermore, the SED of the SN is not consistent with significant reddening given reasonable assumptions about the photospheric temperature (see Figure~\ref{fig:BBpeak}).  We therefore do not correct for extinction in the following analysis.

Our $z$-band PS1 observations of PS1-12sk, tracking the rise of the SN in 4~epochs over $\sim\zrise$~d, are significant because, as noted by  \cite{Pastorello08}, no past SN~Ibn was observed during the rise to maximum.  A possible exception is SN~2011hw, which was observed to rise to over $\sim7$~d after discovery, but only by $\sim0.1$~mag \citep{Smith12}.  This is within the photometric uncertainty in the observations.  Our pre-explosion non-detections constrain the rise time of PS1-12sk to be $9\lesssim\tau\lesssim\zrisemax$~d.  In contrast, pre-detection limits for the Type~Ibn SN~1999cq suggest a very steep rise ($\lesssim4$~d in unfiltered light) \citep{Matheson00}.  For SN~2006jc, pre-explosion limits suggest a rise time $\lesssim17$~d \citep{IAUC8762}, and similarly for SN~2002ao ($\sim20$~d, \citealt{Pastorello08}), while constraining pre-explosion limits are not available for SNe~2000er and 2011hw.  

$BVRI$ photometry for SN~2006jc is also shown in Figure~\ref{fig:LC}.  Because the rise phase of SN~2006jc was not observed, the epoch of peak luminosity is not well constrained.  At discovery, SN~2006jc had observed AB magnitudes of $M_{B,V,R,I}=[\PeakjcB,\PeakjcV,\PeakjcR,\PeakjcI]$~mag (based on photometry in \citealt{Foley07}), $\sim1.5$~mag less luminous than PS1-12sk at peak.  \cite{Pastorello08} compile the optical light curves of SNe~Ibn and report that SN~1999cq was discovered at $M\sim-20$~mag (unfiltered) and SN~2000er was discovered at $M_R\sim-19.5$~mag, $\sim0.5-1$~mag more luminous than PS1-12sk.  Based on the spectroscopic phase of the SNe, \cite{Pastorello08} argue that SN~2002ao and 2006jc may have been similarly luminous at peak (before discovery).  The $r$ and $i$-band luminosity of PS1-12sk is similar to the observed peak of SN~2011hw \citep{Smith12}, although SN~2011hw declined more slowly and was brighter in $B$ by $\sim1$~mag.

The $(g-r)$ color curve of PS1-12sk does not show significant evolution at the precision of our photometry, maintaining $(g-r)\approx 0\pm\gmrmederr$~mag from $\sim-5-+35$~d (Figure~\ref{fig:LC}).  For the purpose of comparison to the PS1-12sk data, we have approximately transformed the $(B-R)$ colors of SN~2006jc and 2011hw to $(g-r)$ via Lupton (2005)\footnote{http://www.sdss3.org/dr9/algorithms/sdssUBVRITransform.php}$^,$\footnote{Using the observed spectrum of PS1-12sk near peak, we estimate that $K$-correction from $z=0.054$ would have a $\lesssim0.1$~mag effect on the $(g-r)$ color, which we neglect here.}.  During the observed rise, the $(g-r)$ color of PS1-12sk reddens only slightly ($\sim0.07$~mag over 7~d; measured only at the $\sim1\sigma$ level in our PS1 photometry).  SN~2006jc evolved from $(g-r)\sim-0.5$~mag at discovery to $(g-r)\sim-0.8$~mag about two weeks later and then returned to $(g-r)\sim-0.5$~mag.  This indicates bluer colors than PS1-12sk, although this level of color evolution 
cannot be ruled out for PS1-12sk given the uncertainty in our photometry after peak.  The color of SN~2011hw was $(g-r)\sim-0.8$~mag, significantly bluer than that of PS1-12sk, and remained 
essentially constant from $\sim15-60$~days \citep{Smith12}.

The NIR-optical SED of PS1-12sk, shown in Figure~\ref{fig:BBpeak}, resembles a single-component power law.  We use a Markov Chain Monte Carlo method \citep{emcee} to fit a blackbody model to the NUV-NIR portion of the SED.  The spectral peak of the SED is not well constrained by our NUV observations, and the NUV spectrum of SNe~Ibn is known to be contaminated by iron emission due to X-ray fluorescence \citep{Foley07}. We therefore adopt a nominal temperature for our blackbody fit by introducing a Gaussian prior of $T=(17\pm2)\times10^3$~K. This is similar to the temperatures near peak of other SNe showing strong circumstellar interaction \citep[see e.g.][ and references therein]{Smith06gy}.  Given this prior, we obtain a radius of $(\peakBBrad\pm\peakBBradD)\times10^{15}$~cm for the photosphere near the epoch of peak brightness.  We place a limit of $\sim10^{-17}~\rm{ergs~s}^{-1}~\rm{\AA}^{-2}$ on any NIR excess above a blackbody SED within the first $1-3$ weeks after maximum brightness.  In 
SN~2006jc, an NIR excess corresponding to an additional warm (dust) blackbody component was detected at the onset of NIR monitoring at $\gtrsim50$~d \citep{DiCarlo08,Smith08}.  In the case of the more luminous SNe~Ibn PS1-12sk  and 2011hw, dust condensation may be delayed due to the higher equilibrium temperatures \citep{Smith12}.  Similarly, dust signatures in the ejecta of SNe~IIn are usually not observed until $\gtrsim200$~d \citep[see e.g.][]{Fassia00,Fox11}.

\subsection{Spectroscopic evolution}
\label{sec:res:spec}

\begin{figure}
\plotone{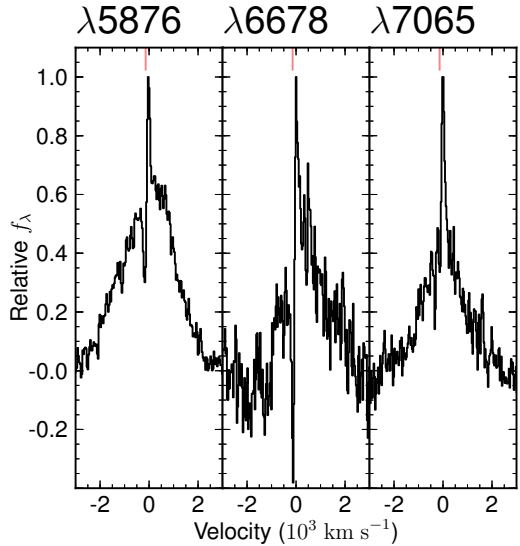}
\caption{\label{fig:specHR}Moderate resolution MMT/BC spectroscopy of \ion{He}{1} features in PS1-12sk from MJD~56014.2, $\specepfour$~d after $z$-band peak.  The velocity scale is centered on the narrow emission component of the P-Cygni profiles, blueshifted by $190~\kms$ from the recession velocity of CGCG~208-042.  The solid red line marks the center of the narrow absorption component, blueshifted by $140~\kms$ relative to the emission.}
\end{figure}

The spectra of PS1-12sk strongly resemble that of the Type~Ibn SN~2006jc at similar epochs (Figure~\ref{fig:spectra}).  The most prominent features in the spectra are the intermediate-width \ion{He}{1} emission lines at $\lambda5876$~\AA{} (Gaussian $\rm{FWHM}\sim3000~\kms$) and $\lambda7065$~\AA{} ($\rm{FWHM}\sim2000~\kms$).  The equivalent width of both lines increases substantially ($EW_{5876}\sim50-100$~\AA) from $\specepzero$ to $\specepone$~days as the continuum fades.  Weaker intermediate-width \ion{He}{1} features at $\lambda\lambda3188,3889,4121,4471,4922,5016,6678,7281$~\AA{} are also visible in the spectra.  

In Figure~\ref{fig:specHR} we display our moderate resolution MMT/BC spectrum of PS1-12sk at $\specepfour$~d.  This spectrum reveals narrow ($\sim100~\kms$) P-Cygni profiles superimposed on the intermediate-width ($\sim3000~\kms$) \ion{He}{1} emission lines.  The narrow emission component is blueshifted from the recession velocity of CGCG~208-042 by $190~\kms$, indicating a velocity offset for the progenitor star that is consistent with the velocity dispersion of CGCG~208-042 ($\sigma=269\pm5~\kms$, from SDSS~DR9 spectroscopy; \citealt{SDSS9}).  Relative to the narrow emission, the P-Cygni absorption features have a blueshift of $\sim140~\kms$ that is consistent between the three strong \ion{He}{1} lines ($\lambda~5876,6678,7065$) visible in this spectrum.

We detect no systematic evolution in the \ion{He}{1} line profiles.  \cite{Smith08} quantify the shape of the line profiles of SN~2006jc as the ratio of the flux summed on the red and blue sides of the \ion{He}{1}~$\lambda\lambda5876,7065$ lines.  Calculating this ratio for the spectra of PS1-12sk, we find Pearson correlation coefficients of $r=\pearsonfiveeightsevensix$ and $\pearsonsevenzerosixfive$, respectively, indicating no statistically significant trends with time.  Significant reddening of the \ion{He}{1} lines in SN~2006jc did not begin until $\sim+50$~days, beyond where our spectral series terminates.  The available spectroscopy does not rule out later dust formation in PS1-12sk, or dust formation enshrouded by optically thick material at higher velocities.

\begin{figure}
\plotone{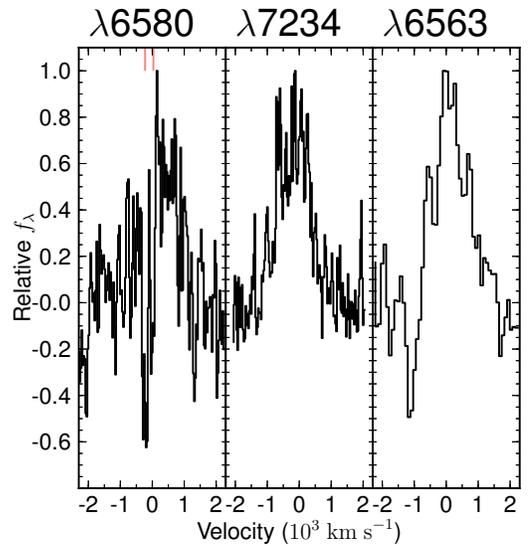}
\caption{\label{fig:specHRC}Left: moderate resolution MMT/BC spectroscopy of \ion{C}{2}~$\lambda\lambda6580,7234$ features in PS1-12sk from $\specepfour$~d after $z$-band peak.  The narrow absorption components corresponding to $\lambda\lambda6578,6583$ are marked with the red solid lines.  Right: The H$\alpha$ feature from the lower-resolution \specepone~days MMT spectrum.  The velocity scale is relative to the \ion{He}{1} narrow emission component, as in Figure~\ref{fig:specHR}.  }
\end{figure}

While H$\alpha$ features are not apparent in our PS1-12sk spectra near peak, intermediate-width H$\alpha$ ($\rm{FWHM}\sim1600~\kms$) emission is present in the $\specepone$~day spectrum (see Figure~\ref{fig:specHRC})).  Similarly, late-developing H$\alpha$ features in the spectra of SNe~2002ao and 2006jc, narrower than those of \ion{He}{1}, was interpreted by \cite{Pastorello07} as evidence for an H-rich circumstellar shell beyond the He shell, which is ionized when the SN flux propagates through the He-rich shell or by the circumstellar interaction.  Significantly stronger H$\alpha$ features were observed in the spectra of SN~2011hw, and discussed by \cite{Smith12} in the context of a CSM with higher H/He abundance than SN~2006jc.  
We revisit the evolution of the \ion{He}{1} and H$\alpha$ features in Section~\ref{sec:WD:H}.

Intermediate-width \ion{C}{2}~$\lambda\lambda6580,7234$ emission is evident in the earliest two MMT/BC spectra ($\specepzero$ and $\specepfour$~days after peak).  \ion{C}{2}~$\lambda6580$ is a blend of lines at $6578$ and $6583$~\AA{}, and in our moderate resolution spectrum we identify narrow absorption components corresponding to both lines at a blueshift consistent with that of the \ion{He}{1} P-Cygni absorption components ($320~\kms$ relative to the velocity of CGCG~208-042; see Figure~\ref{fig:specHRC}).  We note that this intermediate-width \ion{C}{2} emission feature is also visible in the early time (+7 and +13~d) spectra of SN~2011hw (see Figure~3 of \citealt{Smith12}), although there is significant blending with the stronger H$\alpha$ feature in this object.  
Broad photospheric \ion{C}{2} absorption has previously been detected in SN~Ia spectra \citep{Tanaka08,Parrent11}, and a few SNe~Ibc (e.g., SN\,2004aw; \citealt{Taubenberger06}).  The energy levels for these features are high ($16-18$~eV), and \cite{Parrent11} show that the \ion{C}{2} emission is found more often among SNe~Ia with higher effective temperatures.  This temperature dependence is consistent with our detection of \ion{C}{2} lines only in the earlier spectra of PS1-12sk.  
Finally, we do not detect the intermediate-width \ion{O}{1}, \ion{Mg}{2}, and \ion{Ca}{2} emission features that are prominent in the red spectrum of SN~2006jc within $\sim2$~weeks of discovery \citep{Anupama09,Chugai09}.

\section{Host environment properties}
\label{sec:res:host}

PS1-12sk is located 26.5\asec~E and 18.8\asec~N of the giant elliptical galaxy CGCG~208-042, with separation of $27.1\asec$ (projected distance $28.1$~kpc, $\sim1.3$ times the SDSS~DR9 $r$-band Petrosian radius). CGCG~208-042 has redshift $z=0.05402$ (from SDSS~DR9 spectroscopy) and its morphological type is identified as elliptical with 96\% confidence in the Galaxy~Zoo~I catalog \citep{Lintott11}.  \cite{Tempel12} list CGCG~208-042 as the Brightest Cluster Galaxy (BCG) in their friends-of-friends group sample for SDSS~DR8.  The galaxy's host cluster is RXC~J0844.9+4258, listed as a point source in the Northern ROSAT All-Sky X-ray galaxy cluster catalog \citep{NORAS}, with $L_X=0.16\times10^{44}~h^{-2}_{50}~\rm{ergs~s}^{-1}$ ($0.1-2.4$~keV).  

For our adopted value $H_0=71~\kms~\rm{Mpc}^{-1}$, the factor $h^{-2}_{50}\sim0.50$.  Using the best-fit X-ray scaling relation of \cite{Reiprich02}, this corresponds to a total cluster halo mass of $\log(M_{200})=\clusterM~h^{-1}_{50}~\rm{M}_\odot$ ($h^{-1}_{50}\sim0.70$).  This is consistent with the cluster mass derived by \cite{Yang07} from the combined SDSS~DR4 optical luminosity of 15 cluster members: $\log(M)=14.0~h^{-1}~\rm{M}_\odot$.

We produce deep $\grizy$ stacks of PS1 pre-explosion imaging (exposure times of $[45375,47309,118320,151200,75180]$~s, respectively) at the host environment (Figure~\ref{fig:stacks:g:rgb}a).  We use the two-dimensional fitting algorithm \textit{GALFIT} \citep{galfit,galfit3} to model the galaxies within a $1\times1.33$\arcmin\ rectangular region centered on CGCG~208-042 based on the PS1 pre-explosion stacks.  We fit CGCG~208-042 with a Sersic model with Fourier components (to allow for asymmetry) and fit the remaining objects in the field simultaneously as point (PSF models) or extended sources (Sersic profiles; see Figure~\ref{fig:stacks:g:rgb}b).  

\begin{figure}
\plotone{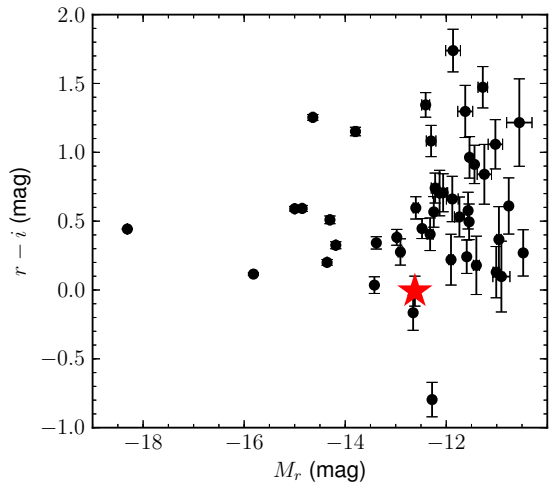}
\caption{\label{fig:CMD} Color-magnitude diagram for sources in the field of CGC~208-042, assumed to be galaxies in the cluster RXC~J0844.9+4258.  The sources were detected on the BCG-subtracted PS1 images displayed in Figure~\ref{fig:stacks:g:rgb}.  The source closest to PS1-12sk is marked with the red star.  CGC~208-042, with $M_r\sim-22.9$~mag and $(r-i)\sim0.5$~mag, is not shown.}
\end{figure}

We subtract the best fit model for CGCG~208-042 and perform object detection and photometry on the BCG-subtracted image (Figure~\ref{fig:stacks:g:rgb}c) using SExtractor.  We exclude objects in a 7\arcsec\ radius from the BCG center, where residuals from the BCG model are strong, and require that sources be detected in the $r$ and $i$ band images.  Assuming all sources in the image are galaxies associated with the cluster, we show a color-magnitude diagram for these \allCdwarfs\ galaxies in Figure~\ref{fig:CMD}.  One galaxy in the field has $M_r=-18.6$~mag, while the rest are dwarf galaxies, with luminosity in the range $M_r\sim[-11,-16]$~mag.  The radial distribution of the dwarf galaxies is similar to the light distribution of the BCG, approximately following a $r^{-2}$ profile.  

The BCG-subtracted PS1 deep stack reveals no source at the position of the supernova PS1-12sk.  By comparison to the faintest objects we detect, this implies a limit of $M_r\gtrsim-10.5$~mag for any unseen, positionally coincident host galaxy.  In our moderate resolution MMT spectrum of PS1-12sk, we do not detect superimposed narrow H$\alpha$ emission with a $3\sigma$ upper limit of $L_{\rm{H}\alpha}\lesssim\MMTHaL~\rm{ergs}~\rm{s}^{-1}~\rm{kpc}^{-2}$ given our $\sim1~\rm{kpc}^2$ aperture.
This limit rules out the association of PS1-12sk with an individual giant HII~region located within a few hundred parsecs, as seen for nearby SN~Ib/c \citep{Crowther12}.
We place a limit on the star formation rate of the local host environment of $\lesssim\MMTsfr~\rm{M}_\odot~\rm{yr}^{-1}~\rm{kpc}^{-2}$ \citep[][Salpeter IMF]{Kennicutt98}, an intensity similar to that of M101 and $\sim3$ orders of magnitude below that of a starburst irregular like NGC~1569 \citep{Crowther12}.  While internal extinction could relax this SFR limit, the color of PS1-12sk is not consistent with significant dust reddening (Section~\ref{sec:res:lc}).

The source nearest to the SN site (separation $\sim2.4$~kpc, assuming it is associated with the cluster) is unresolved in our PS1 images and
has flux $[\gps,\rps,\ips,\zps]=[\galfitblobg,\galfitblobr,\galfitblobi,\galfitblobz]$~mag. Adopting the redshift of the cluster, these magnitudes indicate a source with brightness and color $M_{[\gps,\rps,\ips,\zps]}=[\galfitblobMg,\galfitblobMr,\galfitblobMi,\galfitblobMz]$~mag) similar to blue ultra-compact galaxies found near NGC~4874 in the Coma cluster \citep{Chiboucas11}.  
For comparison, UGC~4904, the host galaxy of SN~2006jc, has $M_g\approx-17$, similar to the Small Magellanic Cloud and $\sim\galfitblobSMCg$~times brighter than that of the faint source near the PS1-12sk explosion site.  SN~2006jc was located near a spiral arm of UGC~4904, visible in SDSS imaging \citep{SDSS9}, at an offset of $\sim1.2$~kpc from the galaxy center.

The dominant source of background flux at the SN explosion site is CGCG~208-042, while no flux is detected from the fainter source at this position.  Based on our \textit{GALFIT} modeling, we estimate that CGCG~208-042 is 3 times brighter at this position than the $3\sigma$ upper limit of flux contributed by other sources.  
We test the hypothesis that the spatial association of PS1-12sk and the putative dwarf galaxy is due to random chance using simulations.  We randomly draw simulated SN positions from a radial distribution consistent with the BCG light over the range $7\arcsec<r<25\arcsec$.  For each simulated SN, we searched for dwarf galaxies within 2\arcsec.  We find that the likelihood of chance association with a dwarf galaxy is $\sim\perblueNeight\%$.  Because the dwarf galaxy near PS1-12sk is among the bluest in the field, with $r-i\sim0$~mag, the likelihood decreases if the association is limited to bluer galaxies.  For  $(r-i)<[0.5,1.0]$~mag, the likelihood is $\sim[\perblueNtwo,\perblueNfour]\%$.  Additionally, we use the method of the SN Legacy Survey \citep{Sullivan06} to order the likelihood of association with potential host galaxy candidates, including the BCG and the nearby faint source described above.  

We calculate the dimensionless $R$ parameter, the elliptical SN-galaxy separation normalized by galaxy size, using the revised formula presented by \cite{Sand11} and the elliptical aperture measured by SExtractor from the $\rps$-band stacked image.  For CGCG~208-042, $R\sim\snlsRrcgc$, while for the fainter host galaxy candidate, $R\sim\snlsRrblob$.  Using PS1 images in other bands (\gps\ips\zps), the $R$ value for CGCG~208-042 is similar and for the fainter candidate is larger ($R\approx7-16$).  We therefore conclude that available observations favor CGCG~208-042 as the most likely host galaxy of PS1-12sk.

While we do not detect evidence for star formation at the PS1-12sk explosion site, there is evidence for centrally-concentrated star formation in CGCG~208-042.  Nebular emission line fluxes are detected in the SDSS $3\asec$ fiber spectrum of CGCG~208-042.  The aperture-corrected star formation rate (SFR) listed in the MPIA/JHU catalog \citep{Brinchmann04} is \MJsfr.  Using the SDSS~DR9 spectroscopy and the emission line ratio diagnostics of \cite{Kewley06}, CGCG~208-042 is classified as a LINER/Composite galaxy, which suggests Active Galactic Nucleus (AGN) contamination of the emission lines which could significantly inflate the SFR estimate \citep{Salim07}. The FUV flux measured by GALEX ($F_{\rm{FUV}}=29\pm4~\mu\rm{Jy}$ in General Release 6; \citealt{GALEX}) provides an independent constraint on the SFR.  Using the \cite{Kennicutt98} scaling relations, the extinction-corrected SFR is $0.33\pm0.03~\rm{M}_\odot~\rm{yr}^{-1}$ (assuming $A_{\rm{FUV}}=0.21$~mag), which is reconcilable with the MPIA/JHU estimate 
given the observational uncertainties, AGN contamination, and the $\sim0.3$~dex scatter observed between FUV and H$\alpha$-based SFR estimates by \cite{Lee09}.  The mass listed for the elliptical galaxy in the MPIA/JHU catalog is \MJmass\, indicating a specific star formation rate of \MJssfr.  

We discuss the possibility of undetected star formation associated with a cluster cool core or dwarf galaxy tidal tail in Section~\ref{sec:prog:young:host}.

\section{Discussion}
\label{sec:disc}

The observations we present in Section~\ref{sec:obs} suggest several key properties for the progenitor star of PS1-12sk.  First, the most prominent features in the optical spectrum are intermediate-width \ion{He}{1} features, reflecting a shock accelerated shell of dense circumstellar medium (CSM) material similar to that accompanying SN~2006jc \citep{Pastorello07,Foley07,Smith08}.  Second, the spectrum shows intermediate-width H$\alpha$ and \ion{C}{2} features indicating additional material within the CSM, as well as unshocked material evidenced by blueshifted narrow absorption features corresponding to \ion{He}{1} and \ion{C}{2}.  Third, the light curve shows a peak magnitude ($\sim-19$~mag) consistent with a luminous SNe~Ib, but declines at a rate similar to SN~1994I and significantly faster than typical SNe Ibc  \citep{Drout11}.  This suggests that the SN has unusual ejecta mass and kinetic energy properties, or that the light curve is powered by a source other than the radioactive decay of $^{56}$Ni.  
Fourth, PS1-12sk was found within a 
galaxy cluster, with no evidence for ongoing star formation at the explosion site.

\subsection{Basic Physical Constraints}

Our $z$-band monitoring of PS1-12sk constrains the rise time to be $\sim9-\zrisemax$~days.  Adopting the rise time as the delay for the forward shock to reach the edge of the He-rich circumstellar material, we use characteristic ejecta velocity to infer the radius of the dense CSM.  While the FWHM of the observed \ion{He}{1} $\lambda5876$~\AA{} line is $\sim3000~\kms$, this may be representative of only the shocked circumstellar material and represents a lower limit for the SN ejecta velocity, $v$.  A velocity, $v\sim10,000~\kms$, is typical of core-collapse SN photosphere velocities \citep{Filippenko97}.  Adopting this value, we find
$R\sim\radRise~\left(\frac{v}{10000~\kms}\right)~\rm{cm}$.
This is consistent with the photosphere radius we infer from the blackbody SED (Section~\ref{sec:res:lc}).  If we further interpret the absorption minimum of the \ion{He}{1} P-Cygni profiles as indicative of the velocity of the mass ejection which populated the He-rich shell, then we infer that the mass ejection event would have taken place within the past $\tEject$~yrs.
We note that our pre-SN PS1 photometry constrains a putative pre-SN transient to $\gtrsim -14$ mag, but the temporal coverage
is not well matched to the expected duration of such an event (see Section~\ref{sec:obs:opt}).

As a basic estimate of the explosion energy, we estimate the integrated bolometric flux of PS1-12sk using our observed photometry.  We interpolate the light curve of PS1-12sk in our most well-sampled band ($z$-band) from $\bolomint$ to $\bolomaxt$~days.  We derive a quasi-bolometric correction for the SN based on photometry from the epoch that is most well-sampled from NUV-NIR (+6~days; Figure~\ref{fig:BBpeak}).  We assume this correction is constant during the duration of our observations, given that we do not find evidence for $g-r$ color evolution (Section~\ref{sec:res:lc}).  Performing Monte Carlo simulations drawing from the uncertainty distribution of our photometry, we find an integrated bolometric energy of $E\gtrsim\boloL\times10^{49}$~ergs.  
If we entirely neglect the NUV ($\lambda<3000$~\AA) flux, the total energy is decreased by only a factor of $\sim2$; this could be more realistic if the high NUV flux we observe near peak is not maintained at later times. 

If we attribute the full bolometric energy output of the SN near peak to interaction with a He-dominated CSM, we can roughly estimate the He ejected in the pre-explosion mass loss.  Using the bolometric energy output estimated for the observed portion of the SN light curve in Section~\ref{sec:prog:LBV:09ip} ($E\sim\boloL\times10^{49}$~ergs) we find that a mass of $M_{He}\sim\boloLHeM~\rm{M}_\odot~ \left({10,000~\kms}/{v_s} \right)^2$ is required assuming full efficiency and pure He composition. This corresponding mass loss rate is $\dot{M}\sim\boloLHeMdot~\rm{M}_\odot~\rm{yr}^{-1}$ given a wind velocity $v_w=100~\kms$ and CSM radius $R=2\times10^{15}$~cm.  Reducing the efficiency increases this $M_{He}$ estimate while decreasing the He fraction decreases it.

\subsection{Radioactive Decay}

If we treat the light curve of PS1-12sk like that of Type~I SNe powered by the radioactive decay of $^{56}$Ni, without contributions from circumstellar interaction, we can estimate the physical parameters of the SN explosion using the SN~Ibc light curve scaling relations of \cite{Drout11}.  These relations are based on the original formalism derived by \cite{Arnett82} and modified by \cite{Valenti08}.  In this framework, our estimates are upper limits and would be reduced by the additional contribution from circumstellar interaction.
Converting the measured peak $\rps$-band magnitude to $R$-band \citep{Windhorst91}, we find $M_R\sim\skMR$~mag (Vega), which corresponds to a nickel mass of $M_{\rm{Ni}}\lesssim\skMNione~M_\odot$.  This upper limit is $\sim2\sigma$ above the population median for SNe~Ib from the \cite{Drout11} sample, $\langle M_{\rm{Ni}}\rangle=0.2\pm0.16~M_\odot$.  \cite{Pastorello08} invoked a range of models with $M_{\rm{Ni}}=0.25-0.40~M_\odot$ to fit the bolometric light curve of SN~2006jc, while \cite{Tominaga08} report an estimate, $M_{\rm{Ni}}=0.22~M_\odot$.  While these masses are larger than those of typical core-collapse SNe, they fall far below the tens of solar masses of $^{56}$Ni produced by models of the most massive exploding stars, e.g. pair instability supernovae \citep{Heger02}.

The measured luminosity and light-curve width for PS1-12sk imply an ejecta mass $M_{\rm{ej}}\sim\skMejone~M_\odot$ and kinetic energy $E_{K,51}\sim\skEKone$ in units of $10^{51}$~ergs.  
From our polynomial fit to the $r$-band light curve, we find $\Delta m_{15}=\PolyPeakrdmfift\pm\PolyPeakrdmfifterr$~mag, which reflects a faster decline than any object in the sample of \cite{Drout11} (except perhaps the fast-fading SN~1994I, $\Delta m_{15}\sim1.4$~mag) and corresponds to a light curve width parameter, $\tau\sim5$~d.  These values are significantly lower than typical SN Ib, which have $\langle M_{\rm{ej}}\rangle=2.0\pm1.0~M_\odot$ and $\langle E_{K,51}\rangle=1.2\pm0.6\times10^{51}$~ergs.  The result that $M_{\rm{Ni}}>M_{\rm{ej}}$ suggests that a process other than radioactive decay, presumably circumstellar interaction, is the dominant power source for the light curve.

\begin{figure}
\plotone{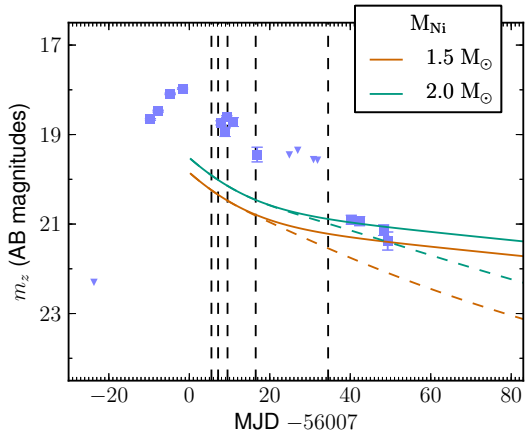}
\caption{\label{fig:Codecay}The $z$-band light curve of PS1-12sk alongside the nebular phase Type~I SN light curve models described in Section~\ref{sec:prog:LBV:model}.  The models are shown with full (solid) and incomplete (dashed) gamma-ray trapping and assuming the explosion date is the same as the discovery date.}
\end{figure}

Using our $z-$band detection of the SN at $+\MNiCoOneT$~days, we report an independent limit on the $^{56}$Ni mass.   To estimate the limit, we fit the analytic Type~I SN nebular phase ($t\gtrsim60$~d) light curve model from \cite{Valenti08} assuming $M_{\rm{ej}}=M_{\rm{Ni}}$, $E_{K,51}=1$, applying a bolometric correction based on the SED observed near peak ($BC=\boloMag$~mag, see Section~\ref{sec:prog:LBV:09ip} and Figure~\ref{fig:BBpeak}), and interpreting the flux observed at this epoch to be wholly powered by the decay of $^{56}\rm{Co}\to^{56}\rm{Fe}$.  
We find $M_{\rm{Ni}}<\MNiCoOne~M_\odot$ if we model incomplete gamma-ray trapping using the exponential gamma-ray absorption probability and homologous density profile adopted by \cite{Valenti08}.  
If we assume full trapping, which may be more realistic when the inner regions of the SN ejecta is dense \citep{Maeda03}, we find $M_{\rm{Ni}}<\MNiCoOneGT~M_\odot$. These models are displayed in Figure~\ref{fig:Codecay}.  
Adopting a more conservative estimate for the rise time, using the date of the latest $z$-band non-detection as the explosion date, has only a small effect on these limits ($M_{\rm{Ni}}<[\MNiCoTwo,\MNiCoTwoGT]~M_\odot$ for [incomplete,full] trapping).  These limits are likely inflated by contributions to the lightcurve from circumstellar interaction, the assumption of Ni-dominated ejecta, and/or if the ejecta is not yet fully nebular at this epoch.

\subsubsection{H Emission}
\label{sec:WD:H}

\begin{figure}
\plotone{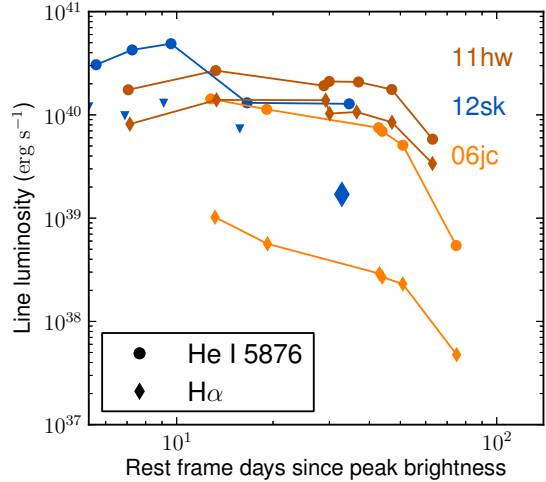}
\caption{\label{fig:lflux}Comparison of intermediate-width line luminosity between PS1-12sk and the SNe~Ibn 2006jc and 2011hw \citep{Smith12}, H$\alpha$ is only detected in the final spectrum (blue diamond) and limits (triangles) are plotted at earlier epochs.}
\end{figure}

The detection of intermediate-width H$\alpha$ emission suggests a significant amount of H remains in the progenitor star near the time of explosion.  To first order, we can estimate $\rm{M_H}$ by starting from our estimate of total CSM mass based on the bolometric energy output, assuming the standard abundance (90\%~He), and attributing the remaining mass to H rather than $\alpha$ elements (Section~\ref{sec:WD:He}).  H would then comprise $3\%$ of the total mass, or $M_H\lesssim\boloLHeMten~\rm{M}_\odot$, however the same order of magnitude uncertainties in velocity, efficiency, and composition apply as in the $\rm{M_{He}}$ calculation.

The H$\alpha$ emission line flux varies by about an order of magnitude in observed SNe~Ibn, both in absolute terms and relative to the strength of the dominant \ion{He}{1}~$\lambda5876$ line.  Figure~\ref{fig:lflux} shows the evolution of these emission lines for well-studied SNe~Ibn.  SN~2006jc had H$\alpha$ emission line fluxes an order of magnitude lower than \ion{He}{1}~$\lambda5876$, but growing relative to He at late times.  SN~2011hw was characterized by much stronger H$\alpha$ emission, only a few times lower than \ion{He}{1}~$\lambda5876$.  In the only epoch where we detect H$\alpha$ in PS1-12sk, it is about an order of magnitude weaker than \ion{He}{1}~$\lambda5876$.  H$\alpha$ emission was not detected in spectra of SN~1999cq \citep{Matheson00}.

\subsection{Host Environment}
\label{disc:host}

The host environments of past SNe~Ibn 1999cq, 2000er, 2002ao, 2006jc, and 2011hw have exclusively been blue, spiral galaxies \citep{Pastorello08,Smith12}.  However, the fact that the first 5 known examples of SN~Ibn occurred in spiral galaxies is not, by itself, strong evidence of a massive star progenitor.  The host environment statistics of SNe~Ia, thermonuclear explosions of white dwarfs, are illustrative.  Only $\sim15\%$ of SN~Ia occur in elliptical galaxies \citep{Li11}, which would suggest a $\sim44\%$ chance that any 5 arbitrary SNe~Ia would be found in spiral hosts.  The host galaxy type statistics of past SNe~Ibn are therefore not sufficient to rule out a low-mass progenitor population like that observed for SNe~Ia.

A massive star progenitor of PS1-12sk carries with it the expectation of significant star formation in its host environment, which is uncharacteristic of the outskirts of massive elliptical galaxies like CGCG~208-042.  Core-collapse SNe are extremely rare in early-type host galaxies, and we know of no core-collapse SNe discovered in association with a BCG.  
In an examination of $2104$ morphologically-classified SN host galaxies in the SDSS field, \cite{Hakobyan12} report only four core-collapse supernovae in S0 or elliptical galaxies.  Core-collapse SN hosting ellipticals often show atypical evidence for star formation via LINER activity and/or galaxy interaction.  In an earlier study, \cite{Hakobyan08} had shown that all other reports of core-collapse supernovae with early type hosts were in fact hosted by mis-classified star forming galaxies.\footnote{The SN~2005md discussed in \cite{Hakobyan08}, reportedly hosted by an elliptical galaxy, was later revealed to be a Galactic cataclysmic varaible star \citep{Leonard10}.}
Similarly, of the SNe discovered by the Lick Observatory Supernova Search (LOSS; \citealt{LOSS1}), $13/536$ SNe~II and Ibc were found in early-type galaxies, all having S0 or S0/a morphology or strong galaxy interaction, or belong to the class of Ca-rich SNe~Ib which may not result from core-collapse in massive stars \citep[see][]{Perets10,Kasliwal12}.  \cite{Suh11} consider a sample of 7 SN~II with early-type host galaxies and find that they systematically show more evidence 
for star formation than SN~Ia hosts, typically being NUV-bright ($\rm{NUV}-r<5.6$~mag) indicating the existence of 
recent star formation.  The most likely host galaxy of PS1-12sk, CGCG~208-042, has GALEX $\rm{NUV}=19.8\pm0.1$~mag and ($\rm{NUV}-r)=5.9\pm0.1$~mag, larger than any of the core-collapse SN host galaxies in \cite{Suh11}.  The  Multi-Epoch Nearby Cluster Survey (MENeaCS) reported one core-collapse SN discovered in a red-sequence cluster galaxy, indicating a rate per unit mass of core-collapse (SN~II) to thermonuclear (SN~Ia) supernovae of $\sim1/5$ in this environment \citep{Graham12,Sand12}.  In an earlier survey of 723 cluster early type galaxies and 1326 field early-types, \cite{Mannucci08} reported a core-collapse SN rate of zero.  These rates can be reconciled due to small number statistics; e.g. the MENeaCS rate is based on the detection of only one SN~II, so the $1\sigma$ statistical uncertainty permits relative rates about an order of magnitude below this.   

Ongoing star formation in the PS1-12sk host environment could be associated with cooling flow filaments like those seen near Abell~1795, Abell~2597, and other nearby clusters \citep[see e.g.][]{Odea08,McDonald09,McDonald11,Tremblay12}.  The archival XMM-Newton observation of the cluster RXC~J0844.9+4258 (Section~\ref{sec:obs:evla}) has insufficient photons to construct a robust temperature profile to determine whether or not the cluster has a cool core.  However, the global properties of the cluster point towards a cool core. The BCG  1.4~GHz radio luminosity (Section~\ref{sec:obs:evla}) nearly meets the threshold established by \cite{Sun09}, $2\times10^{23}~\rm{W~Hz}^{-1}$ at 1.4~GHz, above which all 69~clusters in their sample have cool cores.  Furthermore, the cluster X-ray emission is centered on the BCG to within the spatial resolution of the XMM-Newton observation ($\sim10\asec$ or $\sim10$~kpc; Figure\ref{fig:hostcomposite}).   
Clusters without cool cores typically have separations $\gtrsim50$~kpc between the BCG and X-ray emission peak \citep[see e.g.][]{Hudson10}.  High resolution and sensitive imaging (e.g. Hubble Space Telescope) is required to search for evidence of optical cooling flow filaments near the explosion site of PS1-12sk.

\label{sec:res:prog}

\section{Progenitor Scenarios}
\label{sec:prog}
\label{sec:prog:LBV}

\subsection{Core Collapse Progenitor Models}
\label{sec:prog:LBV:model}

Several lines of evidence, summarized in Section~\ref{sec:intro}, have suggested massive star progenitors for past SN~Ibn events.  Given the strong spectroscopic similarity between PS1-12sk and past SN~Ibn such as SN~2006jc (Section~\ref{sec:obs:spectra}), it is reasonable to assume that their progenitor channel is physically related.  In particular, the spectroscopic detection of intermediate-width He and H features is consistent with the explosion of a massive Wolf-Rayet (WR) or Luminous Blue Variable (LBV)-like star which recently ejected He-rich material into its CSM.
Previously proposed progenitor models for SNe~Ibn call for very massive stars, like the WR and LBV stars we observe undergoing severe mass loss events in the Local Group.  \cite{Pastorello07} discuss outbursts of LBV stars with initial mass $60-100~\rm{M}_\odot$ as analogs to the 2004 pre-explosion brightening of SN~2006jc, but known outbursting LBV stars have H-rich atmospheres inconsistent with the He-dominated CSM inferred for SNe~Ibn.  \cite{Pastorello07,Pastorello08} further a binary progenitor system for SN~2006jc, composed of a $\sim30~M_\odot$ WR star and $50~M_\odot$ LBV.  Binary common envelope interaction provides a possible mechanism for the ejection of a dense, He-rich shell from the progenitor star while providing another source for the LBV-like eruption observed in conjunction with SN~2006jc.  \cite{Foley07} suggested a progenitor star for SN~2006jc which has recently transitioned from a LBV to a WR star. \cite{Smith12} interpret the Type~Ibn SN~2011hw, with relatively strong H$\alpha$ 
emission, as an intermediate example between SN~2006jc and H-rich Type~IIn SNe.  \cite{Smith12} invoke a progenitor belonging to the Ofpe/WN9 class with initial masses $17-100~\rm{M}_\odot$. 

Under the assumption that PS1-12sk had a massive star progenitor, this event represents the first core-collapse supernova residing within a BCG environment.  Given the low rate of SNe~Ibn relative to all core-collapse SNe ($\lesssim1\%$, \citealt{Pastorello08}), PS1-12sk is a statistically unlikely harbinger of massive star explosions in this unique environment.  If we suppose that ongoing star formation in the environment of CGCG~208-042 functions similarly to classical star formation, then we would expect to see of order $10^2$ SNe~II for each SN~Ibn found in galaxy clusters.

Over-representation of SNe~Ibn among core-collapse SNe in red-sequence cluster galaxies could point toward a top-heavy IMF, preference for particular binary evolution scenarios, or overproduction of main sequence stars in a narrow mass range.  Few observational constraints have been placed on the IMF of star formation in galaxy clusters.  \cite{Maoz04} and \cite{Maoz10} have found that the iron enrichment level in galaxy clusters paired with the observed cluster SN~Ia rate may point to a top-heavy IMF.  However, \cite{Mannucci08} find that the rate of SN~II in late type galaxies is independent of environment (cluster or field), which they interpret as evidence for a consistent IMF.  In the cluster core environment, \cite{McDonald11} posed that the FUV/H$\alpha$ excess seen in cooling flow filaments could be interpreted as either evidence for a modest level of dust extinction or a top-heavy IMF.

The association of PS1-12sk with the cluster BCG, rather than another red sequence cluster galaxy, is consistent with a massive progenitor star originating from a cooling flow. 
The global properties of the cluster are consistent with a cool core (Section~\ref{disc:host}).
Narrow-band H$\alpha$ imaging or spatial high-resolution X-ray observations could reveal the presence of a gas filament at the position of the PS1-12sk explosion site, which would be indicative of star formation below the level of our spectroscopic limits.  Additionally, HI radio observations could reveal or constrain the presence of star forming gas near the SN explosion site.

\label{sec:prog:young:host}
The SN could also be associated with the fainter candidate host galaxy discussed in Section~\ref{sec:res:host} if it has an unobserved, low surface brightness tidal tail.  In that case, the likelihood of association we calculate based on the PS1 pre-explosion stacks could be underestimated.  However, such a large offset from a star-forming host galaxy would be unusual for a core-collapse SN.  Among the 36 SNe~Ibs and 35 SNe~IIns in the sample of \cite{Kelly12}, the median offset is $<1$~host galaxy half-light radii and the largest offset among these types is only 2.5~radii.  The offset of PS1-12sk from the faint host galaxy candidate is $3.3$~half-light radii, and this number may be underestimated as the faint source is unresolved.
It is also unlikely that the SN progenitor is a runaway star.  To achieve the $\sim2.4$~kpc separation observed from the faint host galaxy candidate would require a progenitor lifetime after ejection of $\sim10$~Myr given a velocity $\sim200~\kms$ \citep[e.g.][]{Hoogerwerf01}.  A core-collapse SN is unlikely to result from a star that has achieved a separation $\gtrsim100$~pc \citep{Eldridge11}.

\label{sec:prog:LBV:09ip}

Finally, we consider the possibility that the event PS1-12sk is not a SN explosion, but rather a non-destructive eruptive mass loss event from a massive star.  This issue is raised not only by the 2004 flare that preceded SN~2006jc, which was relatively dim at $\sim-14$~mag. but also by the recent ``restlessness'' of SN~2009ip (\citealt{Mauerhan12}, \citealt{Pastorello09ip}, Margutti et al. in prep.).  The progenitor of SN~2009ip is an LBV which has undergone three luminous ($\sim-14$ to $-18$~mag) outbursts in the past 4~yrs \citep{Foley09ip,Mauerhan12}, and is located $\sim4$~kpc ($\sim2$ half light radii, \citealt{Lauberts89}) from the center of its spiral host galaxy, NGC~7259 ($M_B\sim-18$~mag), similar to the remote placement of PS1-12sk (Section~\ref{sec:res:host}).
Given that $<10^{50}$~ergs was needed to power the light curve of PS1-12sk, and assuming a radiative efficiency $\lesssim0.3$ \citep{Falk77}, this leaves open the possibility that there was not sufficient energy to unbind the progenitor star, and we may see additional flares from PS1-12sk in the future.

\subsection{A Degenerate Progenitor?}
\label{sec:prog:Ia}

While the unusual spectroscopic features of PS1-12sk point to the explosion of a massive star, its host environment is dominated by an old stellar population and shows no direct evidence for ongoing star formation.  The environment is reminiscent of those associated with Type~Ia SNe, the thermonuclear explosions of white dwarfs (WDs) which dominate the supernova population within cluster elliptical galaxies including BCGs \citep[see e.g.][]{GalYam02HST,GalYam03,Maoz04,Sharon07,Mannucci08,Li11,Sand11}.  Examples of SNe~Ia exploding within H-rich CSM have been found in SN~2002ic \citep{Hamuy03,Kotak04}, SN~2005gj \citep{Aldering06}, and PTF11kx \citep{Dilday12}.  
Along this line, we consider the possibility that PS1-12sk marks the disruption of a degenerate progenitor system.  Below we discuss constraints on the range of possible white dwarf progenitor models provided by the observational features of PS1-12sk. 

\subsubsection{Spectroscopic properties}

\label{sec:WD:He}
A degenerate progenitor system model for PS1-12sk must be able to eject H, He, and C a few years prior to the SN explosion. To 
explain the significant He, a putative WD progenitor system must contain a He star: either a degenerate WD or non-degenerate He-burning star \citep[see e.g.][]{Benetti02}. 
It is unclear how a He-burning star could produce a significant outburst a few years before explosion since the accretion rate of such a system is typically $10^{-8}$~$M_{\sun}$~yr$^{-1}$.   Moreover, the lifetime of a He-burning star is relatively short (500~Myr), and therefore these systems are expected to be rare in elliptical galaxies.

Merger models for WD-WD systems with high mass ratios (i.e. a CO WD and He WD pair) predict a significant amount of He ejected from the lower-mass companion into the circumbinary environment.  Such a merger model would require fine tuning to match the observed properties of PS1-12sk.  Using smoothed-particle hydrodynamics simulations, \cite{Dan11} find that $\sim3\%$ of the secondary star mass is ejected from the merger of a 0.5 and $1.2~\rm{M}_\odot$ WD system (this fraction is 10\% in similar simulations by \citealt{Guerrero04}).  This enriches the CSM with a He mass lower than the estimate we infer for PS1-12sk, but within an order of magnitude.  In order for the ejected material to reach the $10^{15}$~cm radius inferred for the SN shock, the merger scenario enforces a strict timescale of a few years for the explosion mechanism.

In double-detonation models for the thermonuclear explosion of merging WD binaries, material from the disrupted secondary WD typically accretes unstably onto the primary star on the dynamical timescale of the system ($10^2-10^3$~s; \citealt{Shen12}).  
A delayed detonation on the viscous timescale of the system ($10^4-10^8$~s; e.g. \citealt{Schwab12}) would be required to allow the ejected material to reach the radius observed in SNe~Ibn.  If physically possible, the secondary star of such a system would necessarily fall in the transitional regime where the donor star is intermediate between low mass ($\lesssim0.3~\rm{M}_\odot$) He~WDs expected to produce long-lived AM~CVn or R Coronae Borealis  stars and higher mass ($\gtrsim0.5~\rm{M}_\odot$) CO WDs expected to produce thermonuclear, sub-Chandrasekhar explosions by direct impact mass transfer (see e.g. \citealt{Dan11}).  Additional simulations are needed to determine if a delayed detonation on the viscous timescale is possible, while also ejecting sufficient material into the CSM to produce the He (and also H and C) features observed in SNe~Ibn.

Typical $0.6-1.0~M_\odot$ CO~WD stars have low masses for H ignition ($\sim10^{-5}-10^{-4}~\rm{M}_\odot$; \citealt{Townsley04}).  Nova events could therefore prevent the accretion of significant H on the surface of the white dwarf, as is suggested by the appearance of intermediate-width H features in the ejecta of PS1-12sk.  However, lower mass WDs have higher H ignition masses \citep[see e.g.][]{Shen09H}.  Low mass, $0.3~\rm{M}_\odot$ He~WDs are expected to have $\sim3\times10^{-3}~\rm{M}_\odot$ of H remaining on their surface as they emerge from binary evolution \citep{Driebe98}.  This may be consistent with our observational upper limit (Section~\ref{sec:WD:H}).  
The gradual burning and depletion of this H does not significantly constrain the timescale over which a putative SN~Ibn binary progenitor system must merge.  For $0.3~\rm{M}_\odot$ He~WDs, \cite{Panei07} predict that the H surface mass is depleted by a factor of $\sim2$ after 27~Myr, and is only depleted by a factor of $\sim5$ from the initial value after 6~Gyr. 
Moreover, in a degenerate progenitor model, nova ejecta may contribute to the CSM formation around the SN progenitor.  The detection of H emission features only at late times could be a clue to the geometry of the system.  The emission may only develop once the ejecta reaches an outlying region of H-rich CSM.

R Coronae Borealis (RCB) stars are surrounded by He-rich circumstellar media and their progenitors therefore may share some similarities to the progenitors of SNe~Ibn.  Leading progenitor channels for RCBs are the merger of He and C-O white dwarfs or the final-helium-shell flash of a post-Asymptotic Giant Branch (AGB) central star of a planetary nebula \citep{Clayton11}. 
While we know of no explosion mechanisms for RCB stars, we consider the comparison between their outflows and the inferred circumstellar properties of PS1-12sk.  The $10^{15}$~cm radius we associate with the PS1-12sk shock is $\sim10^2$~times larger than the envelopes of RCB stars ($\sim70~R_\odot$; \citealt{Schoenberner75}), but $\sim10^4$~times smaller than the maximum extent of the diffuse shell of material surrounding RCB ($\sim4$~pc; \citealt{Clayton11}).   RCB stars have a mass loss rate of $\dot{M}\sim10^{-5}~M_\odot~\rm{yr}^{-1}$ \citep{Clayton11}, several orders of magnitude lower than that necessary to populate the CSM of PS1-12sk within the radius of the shocked region. 
With a lower mass transfer rate are AM~CVn stars, consisting of a white dwarf accreting from a degenerate helium donor star, showing evidence for mass loss rates of $\dot{M}\sim10^{-13}-10^{-5}~M_\odot~\rm{yr}^{-1}$ \citep{Iben89,Solheim10} and could possibly produce a short lived ($5-10$~d), low luminosity ($M_V=-15$ to $-18$~mag) thermonuclear SN \citep{Bildsten07,Shen09He}.

\subsubsection{Light Curve Properties}
\label{sec:WD:LC}

When contributions due to circumstellar interaction are removed, the luminosity of PS1-12sk ($M_{\rps}=\PolyPeakrM\pm\PolyPeakrMerr$~mag) and other SNe~Ibn due to radioactive decay of $^{56}$Ni is significantly less than that of typical SNe~Ia ($M_R$ typically $-18.5$ to $-19$~mag \citealt{Li11}, with dust extinction).
A white dwarf progenitor for PS1-12sk may therefore require a sub-Chandrasekhar mass, for which there are several possible explosion mechanisms including runaway He ignition, He shell deflagration, and CO core detonation \citep[see e.g.][]{Bildsten07,Kromer10,Woosley11}.

Thermonuclear explosion models for sub-Chandrasekhar WD-WD merger are typically invoked to explain low-luminosity, sub-energetic SNe~Ia.  However, higher-mass primary stars will produce more $^{56}$Ni and correspondingly brighter explosions, as luminous as $M_V\sim-20$~mag with $\Delta m_{15}\sim1.3$~mag for a $1.4~\rm{M}_\odot$ primary WD \citep{Kromer10}.  A primary star mass sufficient to produce the $M_R\sim-19$~mag peak luminosity and  $\Delta m_{15}>1$~mag light curve width of PS1-12sk calls for a large mass ratio, sufficient to eject significant material from the secondary star during the merger.  However, the contributions of circumstellar interaction to the light curve need to be modeled in more detail.

Alternatively, the ``enshrouded'' accretion induced collapse (AIC) scenario described by \cite{Metzger09} has been proposed to explain SN~2008ha-like events, sub-luminous Type~Ia SNe with fast rise times ($\sim10$~d) and low peak magnitudes ($\sim-14$~mag; \citealt{Foley08ha,Valenti09}).  In this model, a super-Chandrasekhar binary white dwarf pair merges, stripping $\sim0.1~M_\odot$ of potentially He-rich material into a disk surrounding the explosion.  If the circumstellar density is sufficiently high, and a Ni-rich outflow is produced from the accretion disk, a much brighter SN with spectroscopic signatures of circumstellar interaction (like SN~Ibn) could be produced.  However, AIC progenitors would not be expected to have a significant H envelope remaining at the time of explosion and may not be capable of producing the H$\alpha$ emission observed in the spectra of SN~Ibn.  
We note that \cite{Piro12} predict that AIC explosions will produce a radio transient due to 
synchrotron emission from the resulting pulsar wind nebula.  The predicted timescale for the radio transient to reach peak luminosity is a few months with luminosity $\sim10^{28}-\sim10^{29}~\rm{ergs}~\rm{s}^{-1}$ in the $1-10$~GHz range.  This is $1-2$ orders of magnitude below our JVLA luminosity limit for PS1-12sk at a similar epoch (Section~\ref{sec:obs:evla}).

\label{sec:prog:AIC}

\section{CONCLUSIONS}
\label{sec:conc}

We have assembled multi-wavelength observations of the first $\sim2$~months of the supernova PS1-12sk (Figure~\ref{fig:LC}) and its host galaxy cluster, RXC~J0844.9+4258.  While the explosion properties of PS1-12sk are similar to past examples of SNe~Ibn and may be explained by massive-star progenitor models, its unique host environment is suggestive of an origin from an older stellar population.

Spectroscopy of PS1-12sk clearly classifies it as the sixth discovered Type~Ibn SN, with intermediate-width \ion{He}{1} and H$\alpha$ features similar to SN~2006jc.  Our high-resolution spectrum obtained $\sim1$~week after peak brightness shows P-cygni profiles associated with the \ion{He}{1} features (Figure~\ref{fig:specHR}), with absorption minima suggesting a wind velocity of $140~\kms$ for the circumstellar material, and also reveals the presence of narrow \ion{C}{2} absorption.   

The PS1 MDS first detected the SN $\sim9$~d before peak, providing the most detailed constraints to date on the rise of a Type~Ibn SN.  The luminosity of PS1-12sk (with peak $M_{\rps}=\PolyPeakrM\pm\PolyPeakrMerr$~mag) is intermediate between the archetypal Type~Ibn SNe~2006jc and brighter examples like SN~2000er.  The NUV-NIR SED (Figure~\ref{fig:BBpeak}) of the SN within a few weeks of peak brightness is consistent with a single-component blackbody.  Assuming a temperature of $T\sim17\times10^3$~K, we measure a radius of $\sim\peakBBrad\times10^{15}$~cm for the blackbody photosphere, and a similar value based on the rise time and wind velocity.  We do not detect an NIR excess or reddening of \ion{He}{1} line profiles in our optical spectra, suggesting no significant dust formation in the ejecta of PS1-12sk in the first month after peak brightness.  

The host environment of PS1-12sk is unique among Type~Ibn SNe, and is unprecedented among core-collapse SNe.  We use deep imaging from pre-explosion stacks of PS1~MDS observations to characterize the host cluster RXC~J0844.9+4258 (Figure~\ref{fig:stacks:g:rgb}).  We find that the most likely host galaxy for the SN is the BCG of the cluster, CGCG~208-042, at a separation of 28~kpc.  There is no evidence of star formation at the PS1-12sk explosion site, and our spectroscopy yields a limit of $L_{\rm{H}\alpha}\lesssim\MMTHaL~\rm{ergs}~\rm{s}^{-1}~\rm{kpc}^{-2}$.   We identify a nearby faint source which may be a dwarf galaxy and could be responsible for star formation near the host environment, although the separation is several times the source's half-light radius at $\sim2.4$~kpc.  The radio and X-ray characteristics of the host galaxy cluster also suggest a cooling flow may exist that could support star formation in low-luminosity filaments.  

The discovery of PS1-12sk in association with a brightest cluster galaxy either represents surprisingly vigorous star formation, possibly with a top-heavy IMF, in this unique environment or suggests the possibility of a progenitor channel other than the WR or LBV-like, massive star models favored for SN~2006jc.  However, the observational constraints we present for PS1-12sk make it challenging to formulate a degenerate progenitor model capable of producing the He-rich circumstellar medium attributed to SNe~Ibn.  Additional theoretical work is needed to interpret the power source for the light curve of PS1-12sk, the pre-explosion mass loss rate of H and He into the circumstellar medium, and the star formation properties of the explosion environment in terms of a consistent physical progenitor model.  
Additional deep optical imaging observations of the cluster RXC~J0844.9+4258 are needed to search for filamentary emission at the explosion site of PS1-12sk.  Future SN discoveries will provide the sample size necessary to unambiguously associate SNe~Ibn with a stellar population older or younger than typical core-collapse SNe \citep[see e.g.][]{Kelly08,Kelly12,Leloudas11,nesIbcHost}, shedding light on the 
nature of their progenitor stars.

\acknowledgements
\label{sec:ackn}

We thank an anonymous referee for insightful comments and S. Benetti, L. Bildsten, W. Fong, M. Geller, A. Gal-Yam, D. Gifford, A. Hakobyan, P. Mazzali, M. McDonald, E. Pian, and K. Shen for helpful conversations.  We acknowledge useful conversations at a Sky House workshop.

This work was supported by the National Science Foundation through a Graduate Research Fellowship provided to N.E.S.  Support for this work was provided by the David and Lucile Packard Foundation Fellowship for Science and Engineering awarded to A.M.S.  A.S.F. acknowledges support through a NSF STS postdoctoral fellowship.  R.P.K. was supported in part by the National Science Foundation under Grant NSF PHY11-25915 to the Kavli Institute for Theoretical Physics.

The Pan-STARRS1 Surveys (PS1) have been made possible through contributions of the Institute for Astronomy, the University of Hawaii, the Pan-STARRS Project Office, the Max-Planck Society and its participating institutes, the Max Planck Institute for Astronomy, Heidelberg and the Max Planck Institute for Extraterrestrial Physics, Garching, The Johns Hopkins University, Durham University, the University of Edinburgh, Queen's University Belfast, the Harvard-Smithsonian Center for Astrophysics, the Las Cumbres Observatory Global Telescope Network Incorporated, the National Central University of Taiwan, the Space Telescope Science Institute, and the National Aeronautics and Space Administration under Grant No. NNX08AR22G issued through the Planetary Science Division of the NASA Science Mission Directorate.

The National Radio Astronomy Observatory is a facility of the National Science Foundation operated under cooperative agreement by Associated Universities, Inc.  Observations reported here were obtained at the MMT Observatory, a joint facility of the Smithsonian Institution and the University of Arizona.

Partial support for this work was provided by National Science Foundation grant AST-1009749.  The CfA Supernova Program is supported in part by NSF Grant AST-1211196. This publication makes use of data products from the 2MASS Survey, funded by NASA and the US National Science Foundation (NSF). Computations were run on machines in the Harvard Astronomy Computation Facility supported by the Optical and Infrared Astronomy Division of the Harvard-Smithsonian Center for Astrophysics. Other computations were performed on the Harvard Odyssey cluster, supported by the FAS Science Division Research Computing Group at Harvard University. 

This research made use of APLpy, an open-source plotting package for Python hosted at http://aplpy.github.com, and the NASA/IPAC Extragalactic Database (NED) which is operated by the Jet Propulsion Laboratory, California Institute of Technology, under contract with the National Aeronautics and Space Administration.

{\it Facilities:} \facility{CFHT}, \facility{JVLA}, \facility{FLWO:2MASS}, \facility{Hiltner}, \facility{MMT}, \facility{PS1}, \facility{Swift}

\bibliographystyle{fapj}

\end{document}